\documentclass[conference]{IEEEtran}
\pdfoutput=1

\usepackage[]{graphicx}    
\newcommand{\ignore}[1]{}  
\usepackage[dvipsnames, table]{xcolor}
\usepackage{hhline}
\usepackage[labelfont=bf, justification=centering]{caption}
\usepackage{subfigure}
\usepackage{float} 
\usepackage{multirow}
\usepackage{float}
\usepackage{url}
\graphicspath{{images/}}
\usepackage{amsmath}
\usepackage{soul}
\usepackage{comment}
\usepackage{balance}
\usepackage{gensymb}
\usepackage{cite}
\newcolumntype{M}[1]{>{\centering\arraybackslash}m{#1}}
 \usepackage[compact]{titlesec}
\titlespacing{\section}{0pt}{1ex}{0.5ex}
\titlespacing{\subsection}{0pt}{1ex}{0ex}
\titlespacing{\subsubsection}{0pt}{0.5ex}{0ex}
\usepackage{paralist}
\usepackage[left=0.625in,
            right=0.625in,
            top=0.75in,
            bottom=0.9in]{geometry}

\begin{document}


\title{Cross-link Interference Modeling in 6G Millimeter Wave and Terahertz LEO Satellite Communications}

\author{Sergi Aliaga, Vitaly Petrov, Josep M. Jornet\\
Northeastern University, Boston, MA, USA\\
\{aliaga.s, v.petrov, j.jornet\}@northeastern.edu}

\maketitle

\begin{abstract}
One of the important questions when discussing next-generation near-Earth mmWave and terahertz (THz) band satellite communications as an integral part of the 5G-Advanced and 6G landscape is the potential interference-related issues when deploying such systems. While the space-to-ground and ground-to-space interference has been explored in multiple works already, the interference at mmWave and THz cross-links, the links between the satellites themselves, have not been extensively studied yet. However, severe cross-link interference may both challenge the reliability of the data exchange within the constellation, as well as compromise the efficient co-existence of multiple satellite constellations (i.e., by different providers) covering the same or neighboring areas. In this paper, both relevant mathematical models and extensive simulation studies are presented for cross-link mmWave and THz satellite communications. Our results indicate that the cross-link interference in the considered setups is a non-negligible factor that must be further explored and accounted for in the design and deployment of next-general mmWave and THz satellite communication systems.
\end{abstract}

\section{Introduction}
\label{sec:intro}
Seamless integration of high-rate satellite communications into the networking landscape is one of the major innovations on the way from 5G through 5G-Advanced to 6G systems~\cite{Giordani2021}. Particularly, large constellations of low-earth-orbit (LEO) satellites are envisioned as key enablers for ubiquitous global high-rate connectivity~\cite{6g_LEO_commag}. Complementing wireless local area networks and cellular networks, these near-Earth systems thus complete the \emph{``connectivity triad''} providing high-rate and low-latency Internet access for the users on land, sea, and air~\cite{Zhu2022IoT}.

To achieve the prospective 6G-grade performance levels, the use of highly-directional transmissions is envisioned over wide bands in millimeter-wave (mmWave, $\approx$$30$\,GHz--$300$\,GHz, including the Ka band~\cite{pachler21}) and even terahertz (THz, $300$\,GHz--$3$\,THz) frequencies~\cite{OzgurTHzSpace,aliaga2022joint}. Such directional high-rate links may be used for both satellite-to-ground links (Uplink and Downlink) and for inter-satellite links (ISLs), or \emph{cross-links}. For cross-links, optical spectrum is also under active consideration, providing higher rates than even THz channels, but simultaneously challenged by possible pointing errors of extremely narrow (i.e., laser-formed) beams~\cite{Chaudhary2022hybrid,Carlson2022}.

One of the advantages of exploiting mmWave radio for both access and cross-links is the partial reuse of the hardware, thus making the satellite simpler, lighter, and cheaper. While THz radio is a longer-term goal~\cite{Akyildiz_OLD}, the band is already actively exploited in satellite systems for sensing and imaging~\cite{Roy2021Spaceborne,Brown2033Fundamentals}. THz communications are developing rapidly, with recent standardization efforts~\cite{ieee_standard_thz_m} and multi-kilometer-long THz demos~\cite{Sen2022multi}, while the estimated link budget is sufficient for satellite-to-airplane~\cite{joonas_thz_airplane}. Another advantage of mmWave and THz cross-links is that terrestrial limitations (i.e., signal absorption by atmosphere~\cite{Akyildiz_OLD}) are not that profound in space.

An important challenge in developing mmWave and THz satellite communications is related to the interference such solutions may produce to the existing wireless systems, including networks, radars, Earth exploration satellites, etc.~\cite{PengWC2022}. Recently, numerous studies have been presented aiming to model these effects (i.e.,~\cite{RappaportCL2021} and~\cite{Springer2021}, among many others), as well as suggest novel co-existence solutions~\cite{KaragiannidisJSAC2022,Polese2022}. However, considering (tens of) thousands of LEO communication satellites in the near future~\cite{pachler21}, interference with terrestrial systems or satellite-based sensors is not the only challenge. \emph{Particularly, the interference between directional satellite cross-links needs to be carefully accounted for.} The latter is important to guarantee both reliable data exchange within a single satellite constellation, as well as efficient co-existence of several LEO constellations in neighboring orbits.

Recent efforts here primarily comprise simulation-based and solution-centric studies~\cite{Mendoza2017,MayorgaTWC2021,MayorgaGlobecom2011}. To the best of the authors' knowledge, no comprehensive mathematical model has been proposed to date for the interference among directional mmWave/THz satellite cross-links as a function of radio and orbital parameters. Meanwhile, the models for terrestrial mmWave and THz networks~\cite{HeathInterference,Petrov2017interference} are not directly applicable to satellites due to notably different scenario geometry. \emph{We aim to address the abovementioned gap in this article.}

The contributions of this work are summarized as follows:
\begin{compactitem}
\item \emph{A mathematical framework} capable to characterize the interference among directional mmWave and THz satellite cross-links in various deployment configurations. The framework is flexible and can be further tailored to other deployment configurations of interest.

\item \emph{A comprehensive numerical study} illustrating the key trade-offs between the constellation design parameters and the average cross-link interference in different setups.
\end{compactitem}

The paper is organized as follows. Section~\ref{sec:system_model} introduces the system model and the key assumptions. The mathematical framework to estimate the interference among directional mmWave and THz satellite cross-links is detailed in Sec.~\ref{sec:analysis}. In Sec.~\ref{sec:results}, the developed models are numerically elaborated and also cross-verified using extensive system-level simulations. The concluding remarks are drawn in Sec.~\ref{sec:conclusions}.

\begin{figure*}[!t]
    \centering
    \subfigure[High-level deployment illustration]
    {
        \includegraphics[height=0.26\textwidth]{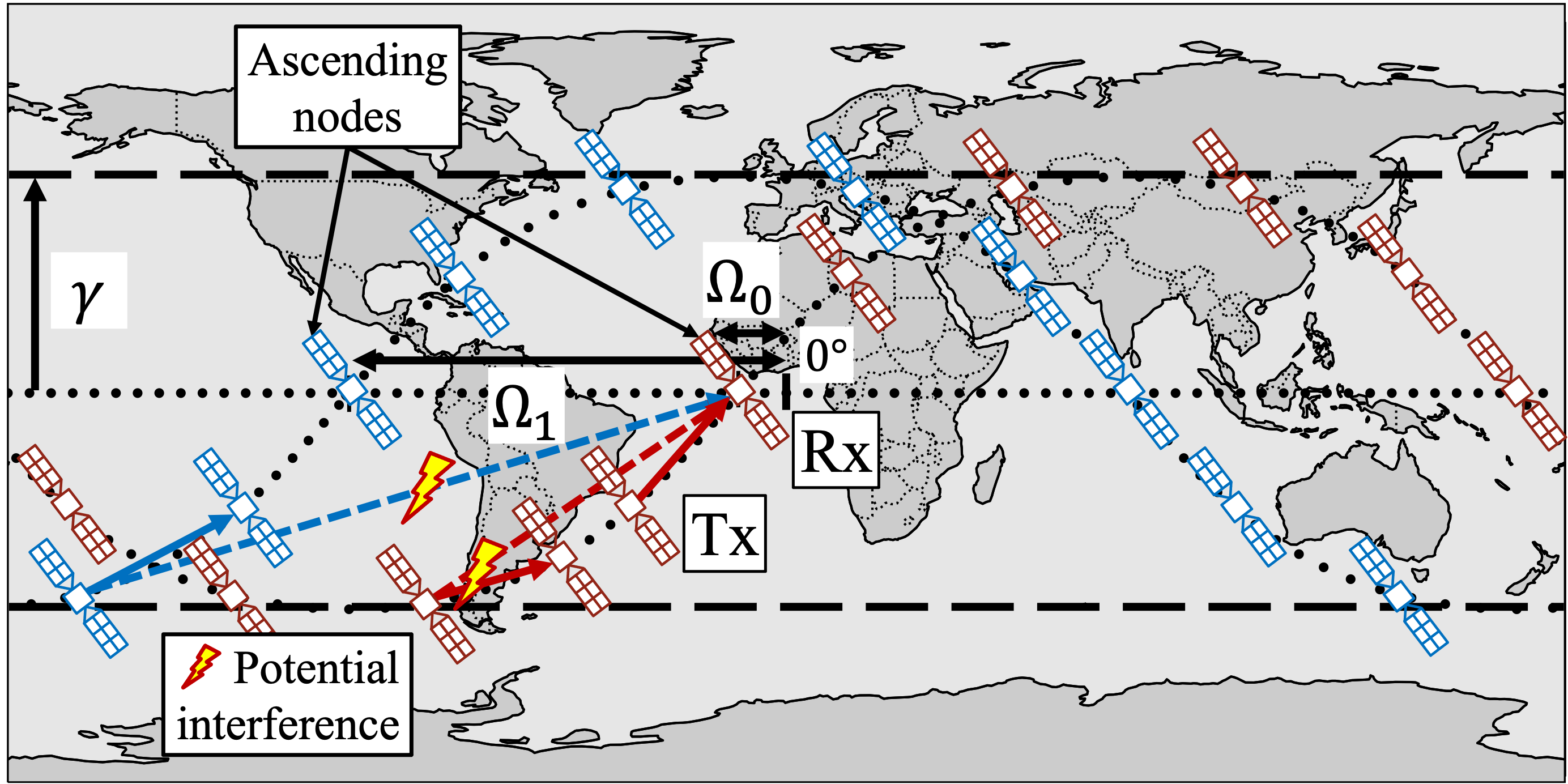}
        \label{fig:orb_2d}
    }
    \subfigure[Key LEO satellite orbital parameters]
    {
        \includegraphics[height=0.26\textwidth]{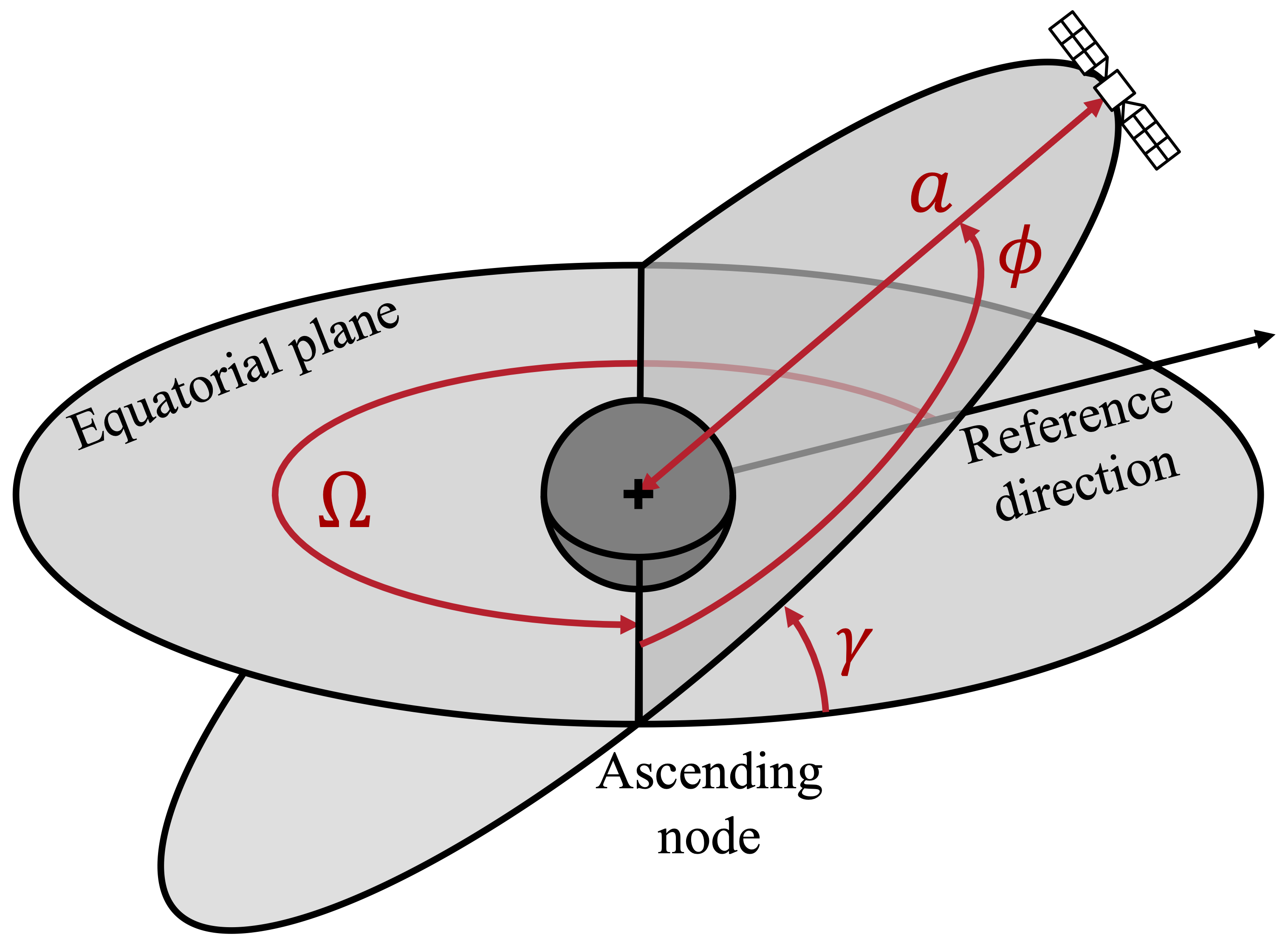}
        \label{fig:orb_parameters}
    }
    \captionsetup{justification=centering}
    \vspace{-3mm}
    \caption{Interference modeling for LEO satellite THz directional cross-link communications.}
    \vspace{-2mm}
    \label{fig:system_model}
\end{figure*}

\section{System Model}
\label{sec:system_model}

\subsubsection{Orbital parameters}
LEO satellites may be deployed in multiple orbits of various configurations. In this work, we target commonly used orbits, as illustrated in Fig.~\ref{fig:orb_2d}, where the satellite placement is mainly determined by the four key orbital parameters, $a$, $\gamma$, $\phi$, and $\Omega$, see Fig.~\ref{fig:orb_parameters}. The first parameter, $a$, is the \emph{semi-major axis}, which for circular LEO orbits is equal to $a = h + R_e$, with $h$ being the orbit altitude above sea level and $R_e$ being the Earth radius. The second one, $\gamma$, is the orbit's \emph{inclination} with respect to the Earth's equator. The third parameter, $\phi$, referred to as \emph{true anomaly}, corresponds to the angle between the satellite position at a certain time and the orbit's \emph{ascending node}, i.e., the crossing point between the orbit and the Earth's equator when going from lower to higher latitudes. The final key orbital parameter, $\Omega$, corresponds to the angular position of the ascending node with respect to a reference direction and is referred to as the \emph{right ascension of the ascending node} (RAAN). Satellites on each of the modeled orbits are placed at equal angular separation from each other.

\subsubsection{Selected scenarios}
We present the interference analysis in directional satellite communications as a decomposed problem, with the aim of understanding the resulting interference levels, the main sources of interference, and the interference pattern across time. For this purpose, we have selected three practical sub-scenarios of increasing complexity, which are the building blocks for the near-future satellite mega-constellations under development:
\begin{compactenum}
    \item \emph{Single orbit}: The simplest option where the interference comes from the satellites located in the same orbit.
    \item \emph{Co-planar orbits}: A more sophisticated case, where interference comes from satellites with different altitude.\footnote{This case is of particular importance to the co-existence in a multi-operator setup, where two constellations have similar parameters, but different altitudes.}
    \item \emph{Shifted orbits}: The most complex for the analysis setup, presenting a typical configuration to obtain a ``shell-like'' pattern to cover a certain range of latitudes~\cite{Leyva2022ngso}.
\end{compactenum}
The three scenarios are depicted in more detail in Sec.~\ref{sec:problem_formulation}.

\subsubsection{Propagation, antenna, and routing assumptions}
Our study focuses on LEO satellites at altitudes from $500$\,km to $2000$\,km, where the effect of atmospheric absorption is already of secondary importance compared to the spreading~\cite{Akyildiz_OLD}. All the satellites are assumed to transmit with the same Tx power, $P_{Tx}$. Hence, the received power, $P_{Rx}$ is primarily determined by the transmit power, antenna gains, and the spreading loss, as $P_{Rx} = \frac{P_{Tx}G^2}{(4\pi d / \lambda)^2}$, where $P_{Tx}$, $G$, and $\lambda$ stand for the transmit power, antenna gain, and the signal wavelength, respectively.

Analyzing the interference in a satellite constellation also requires certain assumptions on the  routing solutions in place. While there can be a wide diversity of options here, it has been assumed that the satellites always transmit to their immediate neighbors in the same orbit, if possible. This is a feasible approach for the discussed deployment and all three modeled scenarios.\footnote{Notably, the contributed model elements can be later recombined in a different manner to account for a specific topology and routing protocol.} We also utilize an analytical \emph{cone-shape} antenna radiation pattern for all the communicating nodes, where the main lobe gain, $G$, is given by $G=2/(1-\cos(\alpha/2))$, where $\alpha$ is the pattern directivity angle/beamwidth~\cite{Petrov2017interference}. The delivered framework can also be further tailored to other radiation patterns, both with and without strong side lobes.

\subsubsection{Metrics of Interest}
In this study, we particularly focus on the average interference power at the target receiving node, $E[I]$, considering the contribution from each potentially interfering satellite, and the signal-to-interference ratio (SIR), $S$, at the receiver, hence putting the interference level in relation with the power of the useful signal from the neighboring satellite. 

\section{Interference Analysis}
\label{sec:analysis}
\label{sec:problem_formulation}
The discussed LEO satellite constellations feature a non-trivial geometry, with complex and even temporarily-dependent mutual locations and orientations of individual satellites. Therefore, for the sake of tractability and clarify of the presentation, we first decompose the considered complex setup into specific smaller elements, particularly modeling the interference coming: (i)~within a single orbit (analyzed in Sec.~\ref{sec:single}), (ii)~from a co-planar orbit (analyzed in Sec.~\ref{sec:coplanar}), and (iii)~from a shifted orbit (analyzed in Sec.~\ref{sec:shifted}). These studied cases allow us to explore directional cross-link interference from different key perspectives.

\subsection{Single orbit}
\label{sec:single}
\begin{figure}[!t]
  \centering
  \includegraphics[width=0.43\textwidth]{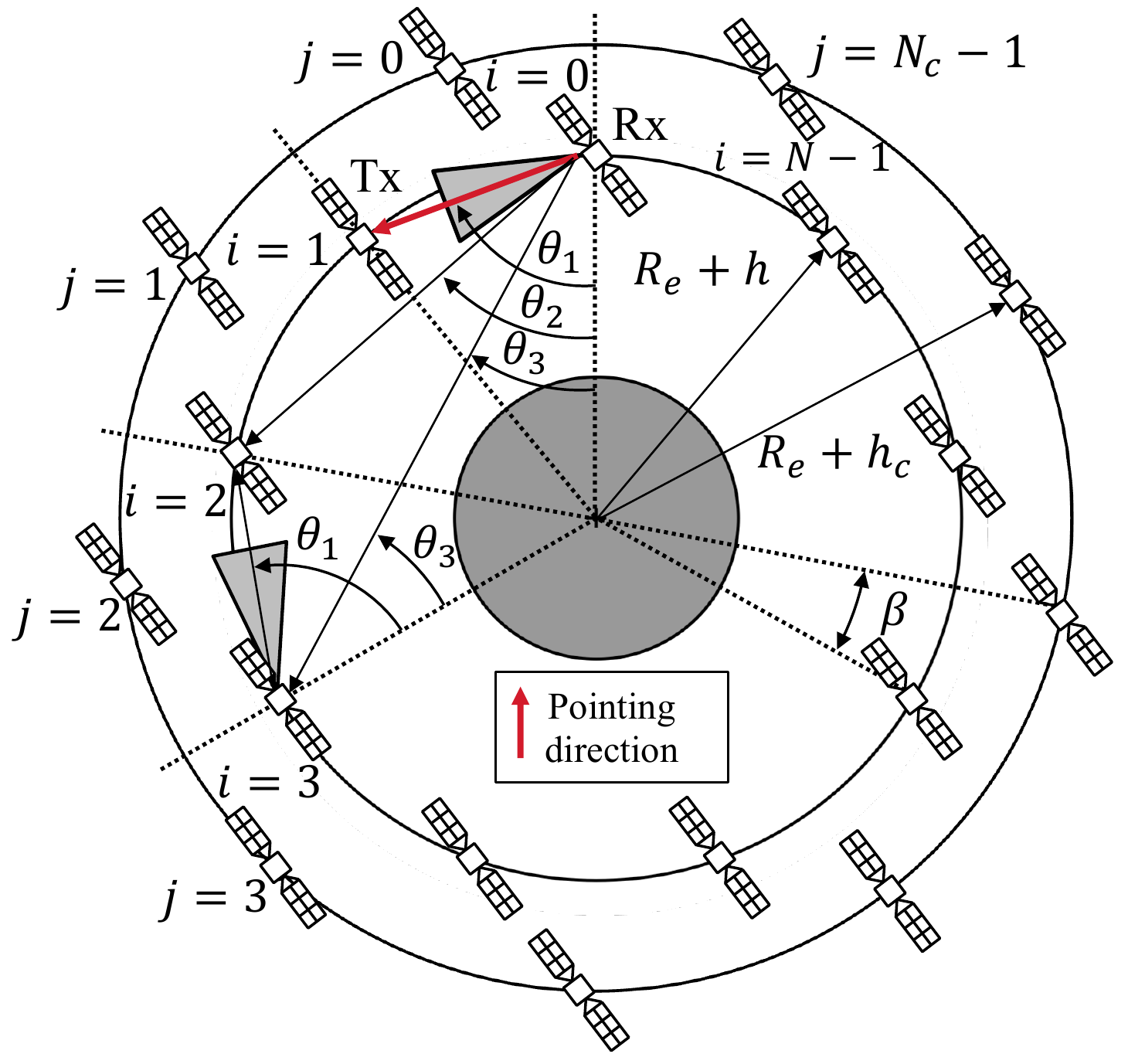}
  \vspace{-2mm}
  \caption{Modeling interference from same or co-planar orbit.}
  \vspace{-3mm}
  \label{fig:single_orbit_geometry}
\end{figure}

Fig.~\ref{fig:single_orbit_geometry} illustrates the geometry for both the single orbit case and the co-planar orbit case analyzed in the following subsection. When the interference, $I_1$, comes from the same orbit, the model comes to $N$ satellites in the same 2D plane forming a circle. We use index $i$ to number those satellites ($i \in [0;N-1]$), where $i=0$ is the target receiver (Rx), $i=1$ is the closest satellite transmitting the useful signal (Tx, see Sec.~\ref{sec:system_model}), while all other satellites are potential interferers.

In the single orbit scenario, the satellite of index $i$ causes interference at the receiver if and only if the following three conditions are simultaneously satisfied: (i)~the interfering signal is not blocked by the Earth; (ii)~the interfering satellite is within the Rx beam; and (iii)~the Rx is within the interferer's beam as well. We analyze those conditions below.

First, the interferer-Rx path is not blocked by the Earth if:
\vspace{-1mm}
\begin{equation}\label{eq:cond1}
    \theta_i > \frac{\pi}{2} - \arccos\left(\frac{R_e}{R_e + h}\right),\ i\in [2, N].
  \end{equation}

Second, based on the scenario symmetry in Fig.~\ref{fig:single_orbit_geometry}, Condition 2, $C2$, and Condition 3, $C3$ are both determined by:
\vspace{-1mm}
\begin{equation}\label{eq:cond23}
    \theta_i > \theta_1 - \frac{\alpha}{2},\ i\in [2, N].
  \end{equation}

Then, considering the satellites are evenly distributed in orbit and applying some basic trigonometry we obtain:
\vspace{-1mm}
\begin{equation}\label{eq:theta_i}
  \theta_i = \frac{\pi}{2} - i\frac{\pi}{N},\ i\in [1, N].
\end{equation}

Substituting (\ref{eq:theta_i}) into (\ref{eq:cond1}) and (\ref{eq:cond23}) thus leads to:
\vspace{-1mm}
\begin{equation}
  i < \frac{N}{\pi} \arccos\left(\frac{R_e}{R_e + h}\right)\enspace{}\land\enspace{}i < 1 + \frac{N}{2\pi}\alpha,\ i\in [2, N].
\end{equation}
Hence, the number of interfering satellites, $N_{i}$ will be the minimum of these two thresholds ($i=1$ is Tx, not interferer):
\vspace{-1mm}
\begin{equation}\label{eq:n_i_single}
  N_i = \lfloor \min(\frac{N}{\pi} \arccos\left(\frac{R_e}{R_e + h}\right), 1 + \frac{N}{2\pi}\alpha) \rfloor - 1.
\end{equation}
Applying the cosine theorem to the geometry in Fig.~\ref{fig:single_orbit_geometry}, we then derive the distance to the interfering satellite $i$, $d_{i}$, as:
\vspace{-1mm}
\begin{equation}\label{eq:single_orb_distance}
  d_i^2 = 2\left(R_e+h\right)^2\left(1-\cos\frac{2\pi}{N}i \right).
\end{equation}
Later, we can combine (\ref{eq:single_orb_distance}) with (\ref{eq:n_i_single}) and derive the average interference power for the single orbit scenario, $E[I_1]$, as:
\vspace{-1mm}
\begin{multline}\label{eq:i1}
  \hspace{-2mm}E[I_1] = \sum_{i=2}^{N_i+1} \frac{\lambda^2 P_{Tx}}{8\pi^2\left(1-\cos\frac{\alpha}{2}\right)^2\left(R_e+h\right)^2\left(1-\cos \frac{2\pi}{N}i\right)}.
\end{multline}
Finally, by calculating the Tx-Rx distance, $d_{1}$, similar to (\ref{eq:single_orb_distance}), we get the average SIR in the single orbit setup, $S_1$, as\footnote{Notably, many input parameters (i.e., the Tx power) cancel in numerator and denominator, so the average SIR depends primarily on the number of satellites, not even their altitude, as further elaborated in Sec.~\ref{sec:results}.}:
\vspace{-1mm}
\begin{equation}\label{eq:s1}
  S_{1} = \frac{1}{\left(1-\cos\frac{2\pi}{N} \right)} / \sum_{i=2}^{N_i+1} \frac{1}{\left(1-\cos\frac{2\pi}{N}i \right)}.
\end{equation}

\subsection{Co-planar orbits}
\label{sec:coplanar}
\begin{figure}[!t]
  \centering
  \includegraphics[width=0.38\textwidth]{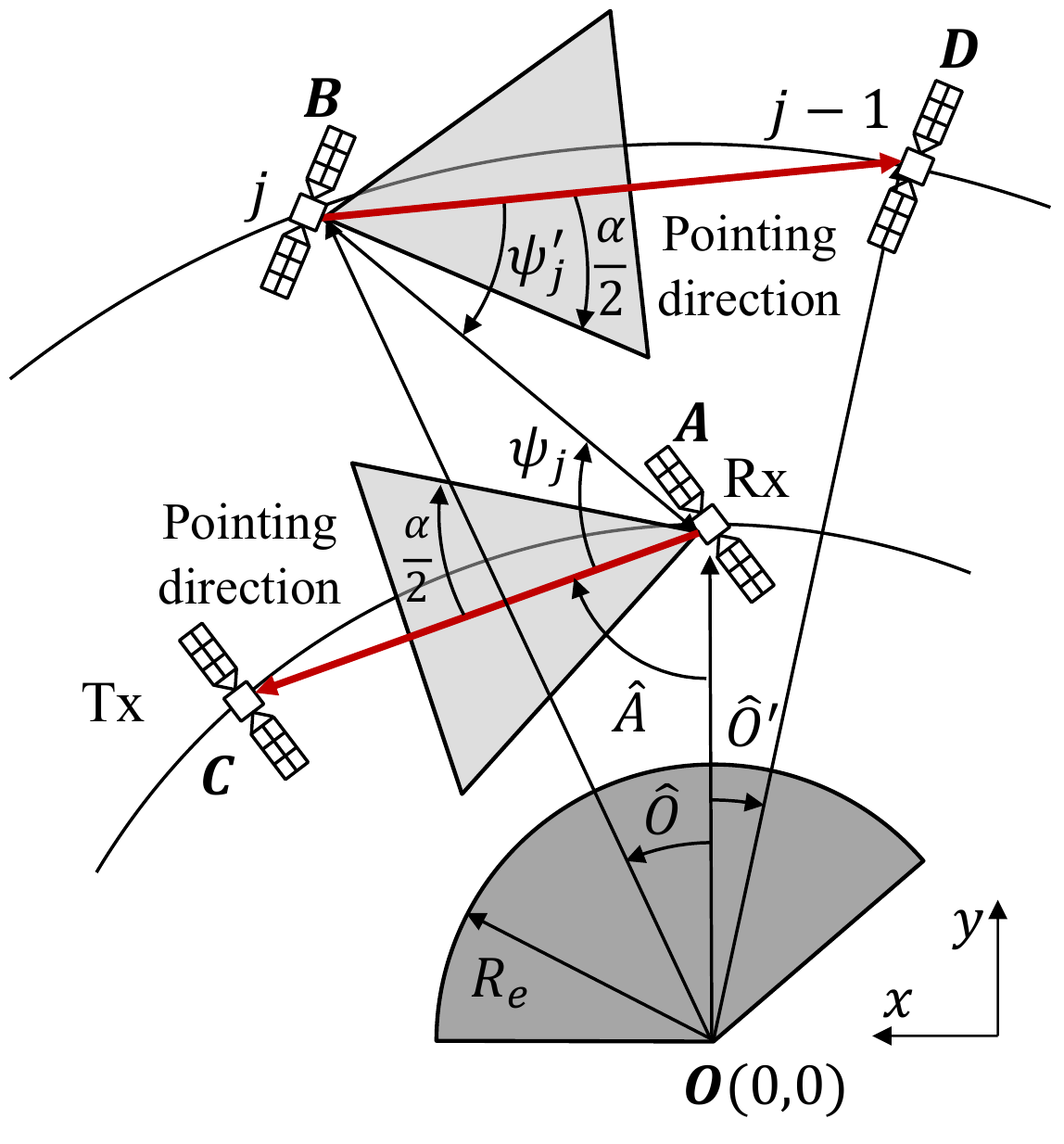}
  \vspace{-1mm}
  \caption{Interference from a co-planar orbit (zoomed).}
      \label{fig:coplanar_orbits_zoom}
    \vspace{-5mm}
\end{figure}

We now proceed with the co-planar orbits setup (see Fig.~\ref{fig:single_orbit_geometry}), illustrating, i.e., the co-existence of two satellite constellations by different service providers. We particularly detail the analysis for a case where the target Rx is at a lower altitude, $h$, and the interference, $I_2$, comes from a higher orbit, $h_c$, while the derivations for the opposite case are obtained similarly. An illustration of the scenario geometry is given in Fig.~\ref{fig:coplanar_orbits_zoom} with $N$ satellites on the lower orbit and $N_{c}$ satellites on the higher orbit. We introduce Cartesian axes $x$ and $y$ to simplify the further derivations by using vector notation.

As the two orbits have different altitudes, they also have different orbital speeds. Hence, there is a new time-dependent parameter $\Delta\beta$, referred to as \emph{relative angular offset}, with the time period determined by Kepler's third law~\cite{Maral2009satellite}.
Similar to the single orbit setup, the three conditions from the previous subsection have to be met for the satellite $j$ to interfere with the Rx. However, as for co-planar orbits $h \neq h_c$, $C2$ and $C3$ are now fulfilled separately.

Specifically, non-blocking by the Earth, $C1$, is satisfied if:
\vspace{-1mm}
\begin{equation}
  \psi_j + \hat{A} > \frac{\pi}{2} - \arccos\left(\frac{R_e}{R_e + h}\right).
\end{equation}

In their turn, mutual antenna alignment conditions, $C2$ and $C3$, are satisfied by $|\psi_j| < \frac{\alpha}{2}$ and $| \psi_j' | < \frac{\alpha}{2}$, respectively.

Using the vector notation, we then derive the angle $\psi_j$ as:
\vspace{-1mm}
\begin{equation}\label{eq:psi_j}
  \psi_j = \arccos \left( \langle \vec{AB}, \vec{AC} \rangle / \| \vec{AB} \| \right),
\end{equation}
where
\vspace{-1mm}
\begin{align}
  &\vec{A} = (0 ,\quad{} R_e + h),\ \vec{B} = (R_e + h_c) (\sin\hat{O} ,\quad{} \cos\hat{O}) \label{vecB},\\
  &\vec{AB} = \vec{B}\hspace{-0.5mm}-\hspace{-0.5mm}\vec{A} = ((R_e + h_c)\sin\hat{O} ,\quad{} (h_c-h)\cos\hat{O}),\\
  &\vec{AC} = (\sin\hat{A} ,\quad{} -\cos\hat{A}),\quad{}\hat{O} = \Delta\beta + 2\pi j / N_c.
\end{align}

The angle $\hat{A}$ corresponds to the angle $\theta_1$ from the single orbit scenario as $\hat{A} = \frac{\pi}{2} - \frac{\pi}{N}$.

Hence, (\ref{eq:psi_j}) can be expanded into
\vspace{-1mm}
\begin{equation}
\psi_j\hspace{-0.5mm}=\hspace{-0.5mm}\arccos\left(\frac{(R_e\hspace{-1mm}+\hspace{-1mm}h_c)\sin\hat{O}\sin\hat{A}\hspace{-1mm}-\hspace{-1mm}(h_c\hspace{-1mm}-\hspace{-1mm}h)\cos\hat{O}\cos\hat{A}}{\sqrt{(R_e\hspace{-1mm}+\hspace{-1mm}h_c)^2\sin\hat{O}^2\hspace{-1mm}-\hspace{-1mm}(h_c\hspace{-1mm}-\hspace{-1mm}h)^2\cos\hat{O}^2}}\right).
\end{equation}

The final key angle here, $\psi_j'$, is obtained similarly to $\psi_j$ as:
\vspace{-1mm}
\begin{equation}
  \psi_j' = \arccos\left( \langle \vec{BD}, \vec{BA} \rangle / \left[ \| \vec{BD} \| \cdot \| \vec{BA}\| \right] \right),
\end{equation}
where
\vspace{-1mm}
\begin{align}
  &\vec{BD} = \vec{D} - \vec{B},\quad{}\quad{}\vec{BA} = \vec{A} - \vec{B},\\
  &\vec{D} = (R_e + h_c)(\sin\hat{O}',\quad{} \cos\hat{O}'),\\
  &\hat{O}' = \Delta\beta + \frac{2\pi}{N_c} ((j-1)\mod N_c-1).
\end{align}

Once the main angles are specified, we can derive the set of interfering satellites, $C$, as $C = C1 \cap C2 \cap C3$, where:
\vspace{-1mm}
\begin{align}
  C_1 &= \{j \in [0,N_c-1]\ |\ \psi_j > \frac{\pi}{N} -\arccos\left(\frac{R_e}{R_e + h}\right) \},\nonumber\\
  C_2 &= \{j \in [0,N_c-1]\ |\ |\psi_j| < \frac{\alpha}{2} \},\nonumber\\
  C_3 &= \{j \in [0,N_c-1]\ |\ |\psi_j'| < \frac{\alpha}{2} \}.
\end{align}

We now derive the distance from the potential interferer $i$ to the target Rx. From the analysis above and the geometry in Fig.~\ref{fig:coplanar_orbits_zoom}, this distance corresponds to $\| \vec{AB} \|$, thus:
\vspace{-1mm}
\begin{equation}
  d_i^2 = (R_e+h_c)^2\sin\hat{O}^2+(h_c-h)^2\cos\hat{O}^2.
\end{equation}

Finally, the average interference power, $E[I_2]$, and the average SIR, $S_{2}$, for the co-planar scenario are derived as:
\vspace{-1mm}
\begin{align}\label{eq:i2}
  E[I_{2}] &= \sum_{j \in C_1\cap C_2 \cap C_3}  \lambda^2 P_{Tx} / [4\pi^2 \left(1-\cos\frac{\alpha}{2}\right)^2 \nonumber \\
   &((R_e+h_c)^2 \sin^2\hat{O}_j) + (h_c-h)^2\cos^2\hat{O}_j)],
\end{align}
\begin{align}\label{eq:s2}
  S_{2} = \frac{1}{2Z\left(1-\cos\frac{2\pi}{N} \right)},
\end{align}
where
\vspace{-1mm}
\begin{align}
\hspace{-2mm}Z\hspace{-1mm}=\hspace{-4mm}\sum_{j \in C_1\cap C_2 \cap C_3} \frac{1}{(R_e+h_c)^2 \sin\hat{O}^2 + (h_c-h)^2\cos\hat{O}^2}.
\end{align}

\vspace{1mm}
\subsection{Shifted orbits}
\label{sec:shifted}
\begin{figure}[!h]
  \centering
  \vspace{-1mm}
  \includegraphics[width=0.38\textwidth]{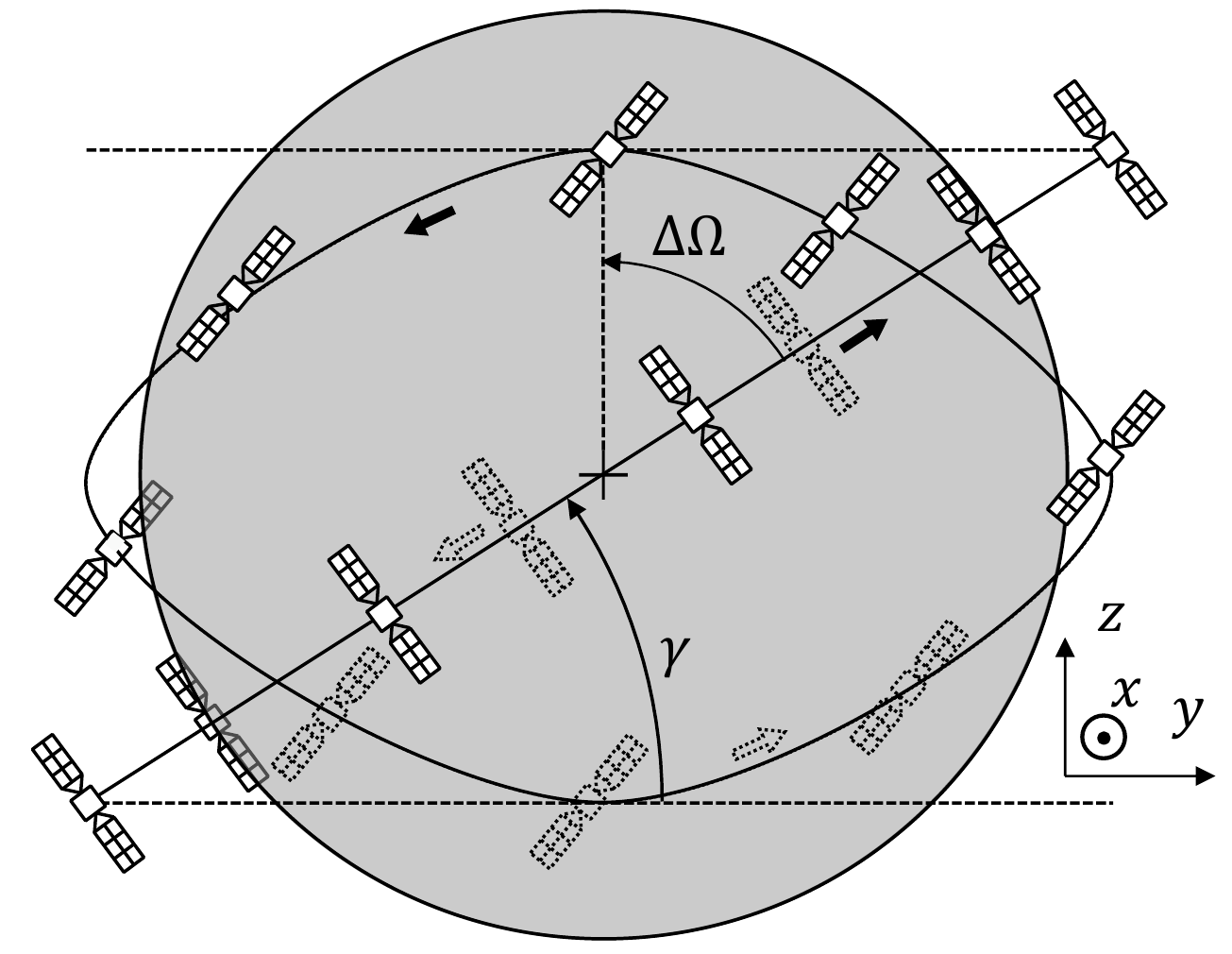}
  \vspace{-1mm}
  \caption{Modeling interference from a shifted orbit.}
  \label{fig:shifted_orbits_geometry}
\end{figure}

The third component of the model covers the case where the interference, $I_3$, comes from an orbit with the same altitude and the same inclination, but a different $\Omega$~\cite{Leyva2022ngso}, as in Fig.~\ref{fig:shifted_orbits_geometry}.\footnote{Many practical constellations employ multiple \emph{shifted orbits} to grant continuous coverage between the latitude range $[-\gamma, \gamma]$, as in Fig.~\ref{fig:orb_2d}~\cite{Leyva2022ngso}.} The angular shift between the orbits is termed as $\Delta \Omega$.

As the satellites in shifted orbits are not always within the same 2D plane, we have to use a 3D Geocentric Equatorial Coordinate (GEC) system fixed with respect to the background stars~\cite{Maral2009satellite}. The $z$-axis corresponds to the Earth's rotation axis, the $x$-axis corresponds to the direction of the vernal equinox, and the $y$-axis follows a right-handed convention with the other two. With this coordinate system, we can identify every satellite's position and pointing directions as 3-element vectors, thus also determining the angles between them.

For a given satellite~$j$ out of $N_s$ in orbit 2 (Orbit 1 contains the target Rx), we begin by identifying its position inside its own orbital plane. Concretely, we first compute its true anomaly (angular position inside the orbital plane) as:
\vspace{-1mm}
\begin{equation}
  \phi_j = \Delta\beta + \frac{2\pi}{N_s}j,\ \forall j \in [0,N_s-1],
\end{equation}
where $N_s$ is the number of satellites in orbit 2 and the relative angular offset $\Delta\beta$ is constant through time since both orbits have the same altitude and, therefore, the same orbital period.

We then get the satellite's position in the orbital plane as:
\vspace{-1mm}
\begin{equation}
  \vec{r}_{0j} = (x_{0j}, y_{0j}, z_{0j}) = (R_e+h)(\cos \phi_j, \sin \phi_j, 0).
\end{equation}

Next, we define the conversion matrix from the orbital coordinate system to the GEC system as:
\vspace{-1mm}
\begin{equation}
  M_k = \begin{bmatrix}
    \cos\Omega_k & -\sin\Omega_k\cos \gamma & \sin\Omega_k\sin \gamma \\
    \sin\Omega_k & \cos\Omega_k\cos \gamma & -\cos\Omega_k\sin \gamma \\
    0 & \sin \gamma & \cos \gamma \\
  \end{bmatrix}
\end{equation}
where $k=0$ corresponds to the orbit with Rx and $k=1$ to the shifted orbit, where the interference comes from. $\Omega_k$ is the orbit's RAAN, and $\gamma$ is the orbit's inclination.
We then obtain the $j$th satellite position in orbit $k$ in the GEC system as $\vec{r}_j = M_k\vec{r}_{0j}$. We use the same procedure to find the transmitter's and receiver's position vectors, $\vec{r}_{Tx}$ and $\vec{r}_{Rx}$, respectively, and validate the interference conditions.

To verify no blockage by the Earth, we get the rise and set between the candidate satellite $\vec{r}_j$ and the receiver $\vec{r}_{Rx}$ as~\cite{Sharaf2012satellite}:
\begin{align}
  \begin{split}
  R_j = &\langle \vec{r}_j, \vec{r}_{Rx}\rangle^2 - \|\vec{r}_j\|^2\|\vec{r}_{Rx}\|^2 + \\
  &+(\|\vec{r}_j\|^2+\|\vec{r}_{Rx}\|^2)R_e^2 - 2R_e^2\langle \vec{r}_j, \vec{r}_{Rx}\rangle.
  \end{split}
\end{align}
and thus establish the set of visible satellites as:
\begin{equation}
  C_1 = \{j \in [0,N_s-1]\ |\ R_j \leq 0 \}.
\end{equation}

To verify $C2$ and $C3$, we compute the distance vector between the Rx and the interfering satellite, and their pointing directions, respectively, as:
\begin{align}
  \vec{r}_{Rx\rightarrow j} &= \vec{r}_{j}-\vec{r}_{Rx},\\
  \vec{r}_{Rx\rightarrow Tx}&= \vec{r}_{Tx}-\vec{r}_{Rx},\\
  \vec{r}_{j\rightarrow j-1} &= \vec{r}_{j-1}-\vec{r}_{j}.
\end{align}
and compute angles $\psi_j$ and $\psi_j'$ as:
\begin{align}
  \psi_j &= \arccos \frac{\langle \vec{r}_{Rx\rightarrow j}, \vec{r}_{Rx\rightarrow Tx} \rangle}{\| \vec{r}_{Rx\rightarrow j} \| \cdot \| \vec{r}_{Rx\rightarrow Tx}\|},\\
  \psi_j' &= \arccos \frac{\langle \vec{r}_{j\rightarrow j-1}, \vec{r}_{j\rightarrow Rx} \rangle}{\| \vec{r}_{j\rightarrow j-1} \| \cdot \| \vec{r}_{j\rightarrow Rx}\|},
\end{align}
where $\vec{r}_{j\rightarrow Rx}=-\vec{r}_{Rx\rightarrow j}$.

Similar to the previous subsection, the set of satellites satisfying $C2$ and $C3$ will then be:
\begin{align}
  C_2 &= \{j \in [0,N_s-1]\mod(N_s)\ |\ |\psi_j| < \frac{\alpha}{2} \},\\
  C_3 &= \{j \in [0,N_s-1]\mod(N_s)\ |\ |\psi_j'| < \frac{\alpha}{2} \}.
\end{align}

These derivations finally lead to the equations for the average interference power, $E[I_3]$, and average SIR, $S_{3}$, for the case where the interference comes from a shifted orbit:
\begin{equation}\label{eq:i3}
  E[I_{3}] = \sum_{j \in C_1\cap C_2 \cap C_3} \\ 
  \frac{\lambda^2 P_{Tx}}{4\pi^2 \left(1-\cos\frac{\alpha}{2}\right)^2 \|\vec{r}_{Rx\rightarrow j}\|^2},
\end{equation}
\begin{equation}\label{eq:s3}
  S_{3} = 1 / \left( 2(R_e+h)^2\left(1-\cos\frac{2\pi}{N} \right) \sum_{j=0}^{N_j} \frac{1}{\|\vec{r}_{Rx\rightarrow j}\|^2} \right).
\end{equation}

\section{Numerical Results}\label{sec:results}
In this section, the mathematical models developed above get numerically elaborated. We also complement the analytical results with those produced within an extensive system-lever simulation campaign. For this purpose, the in-house developed satellite communications simulator from~\cite{Alqaraghuli2021compact} has been utilized. The framework is implemented in Python, it accurately models massive satellite constellations with communications and sensing capabilities deployed in most of the practical orbits~\cite{aliaga2022joint}.

\begin{figure}[!t]
   \centering
   \includegraphics[width=0.95\columnwidth]{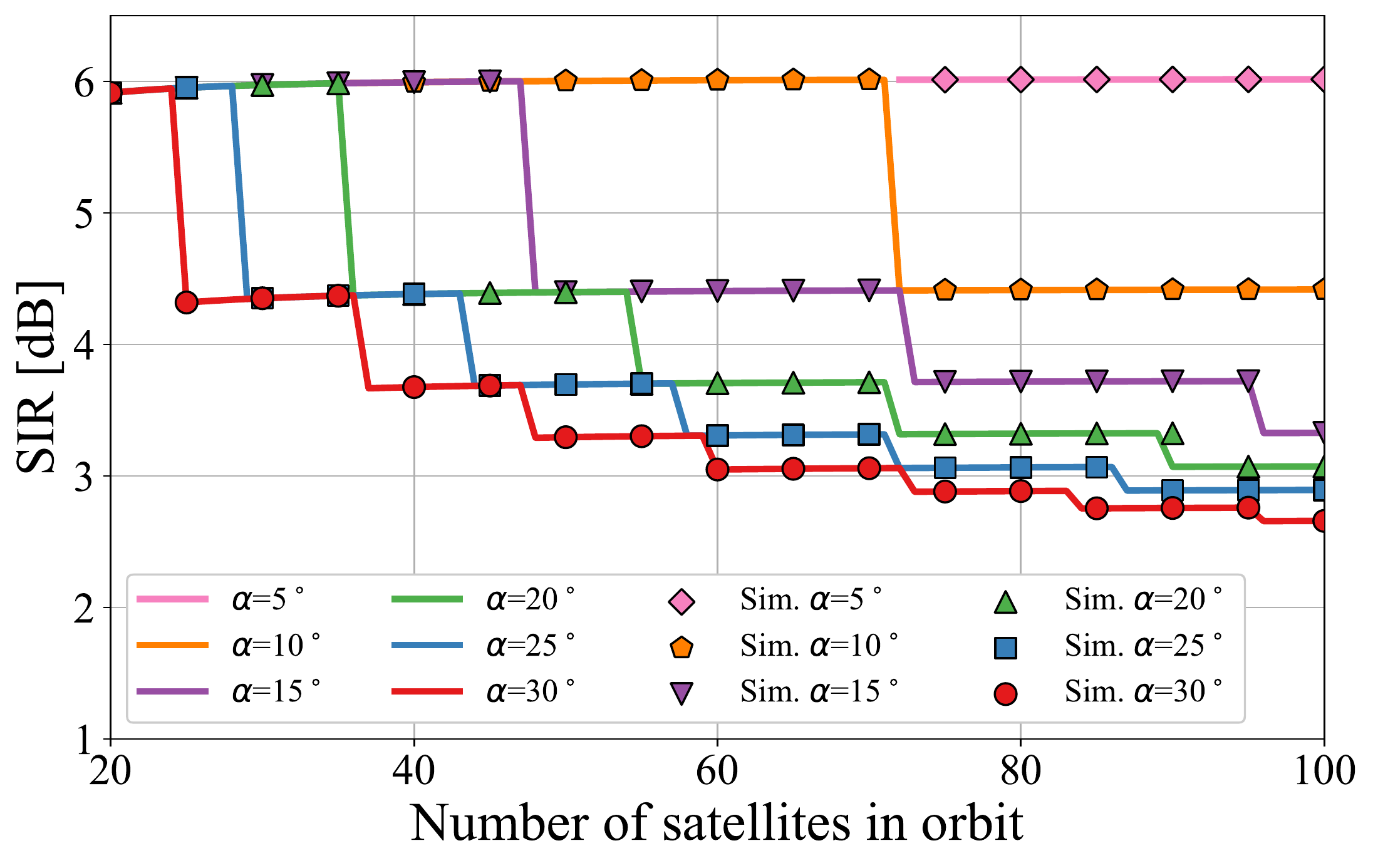}
   \vspace{-2mm}
   \caption{Average SIR with interference from the same orbit.}
   \vspace{-3mm}
   \label{fig:single_orbit_results}
\end{figure}

\begin{figure*}[!t]
    \centering
    \subfigure[SIR evolution in time for an orbit separation of $10$\,Km]
    {
        \includegraphics[height=0.29\textwidth]{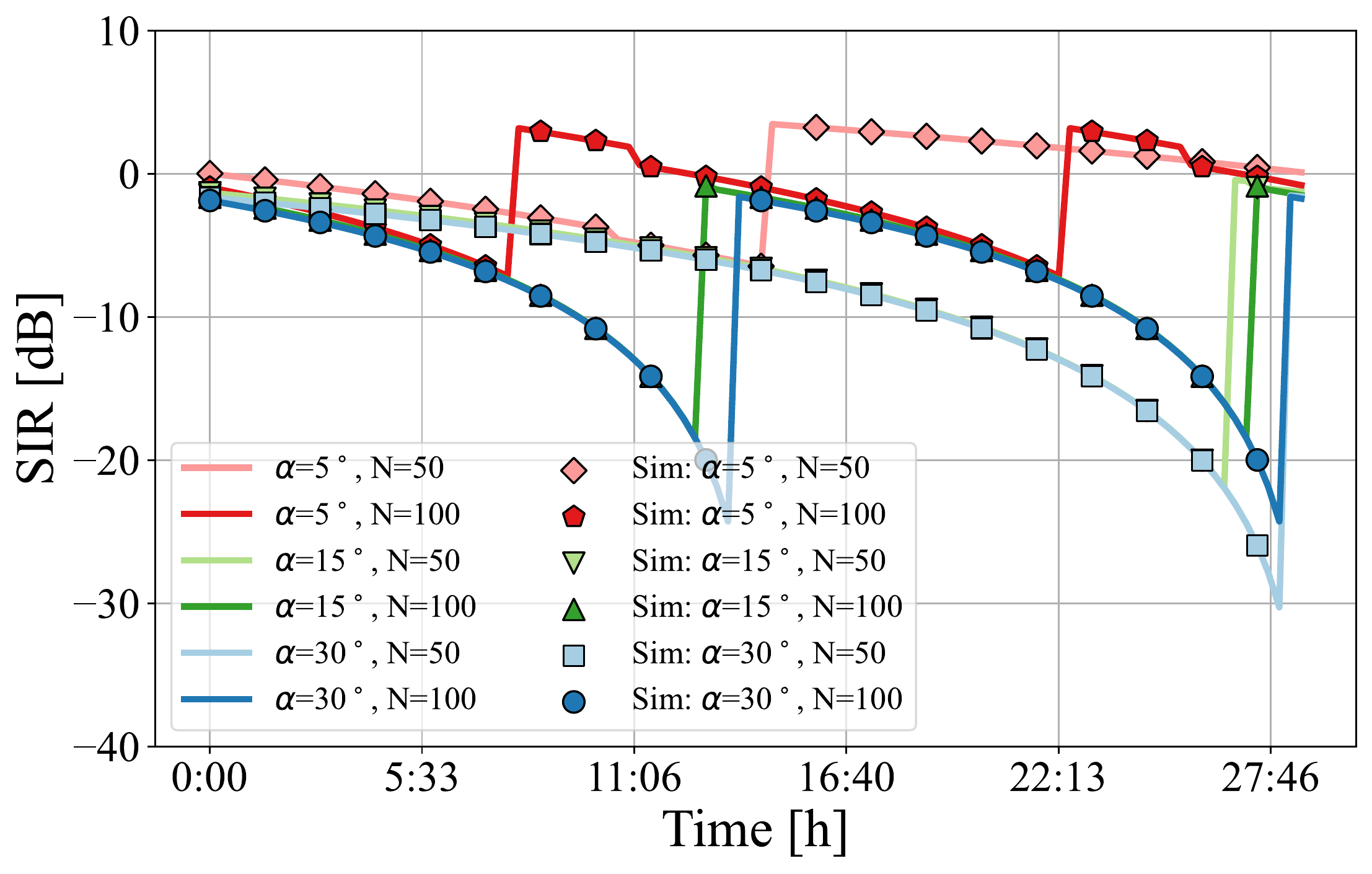}
        \label{fig:coplanar_SIR_time_results}
    }
    \hfill
    \subfigure[Expected interference for different orbit separation]
    {
        \includegraphics[height=0.29\textwidth]{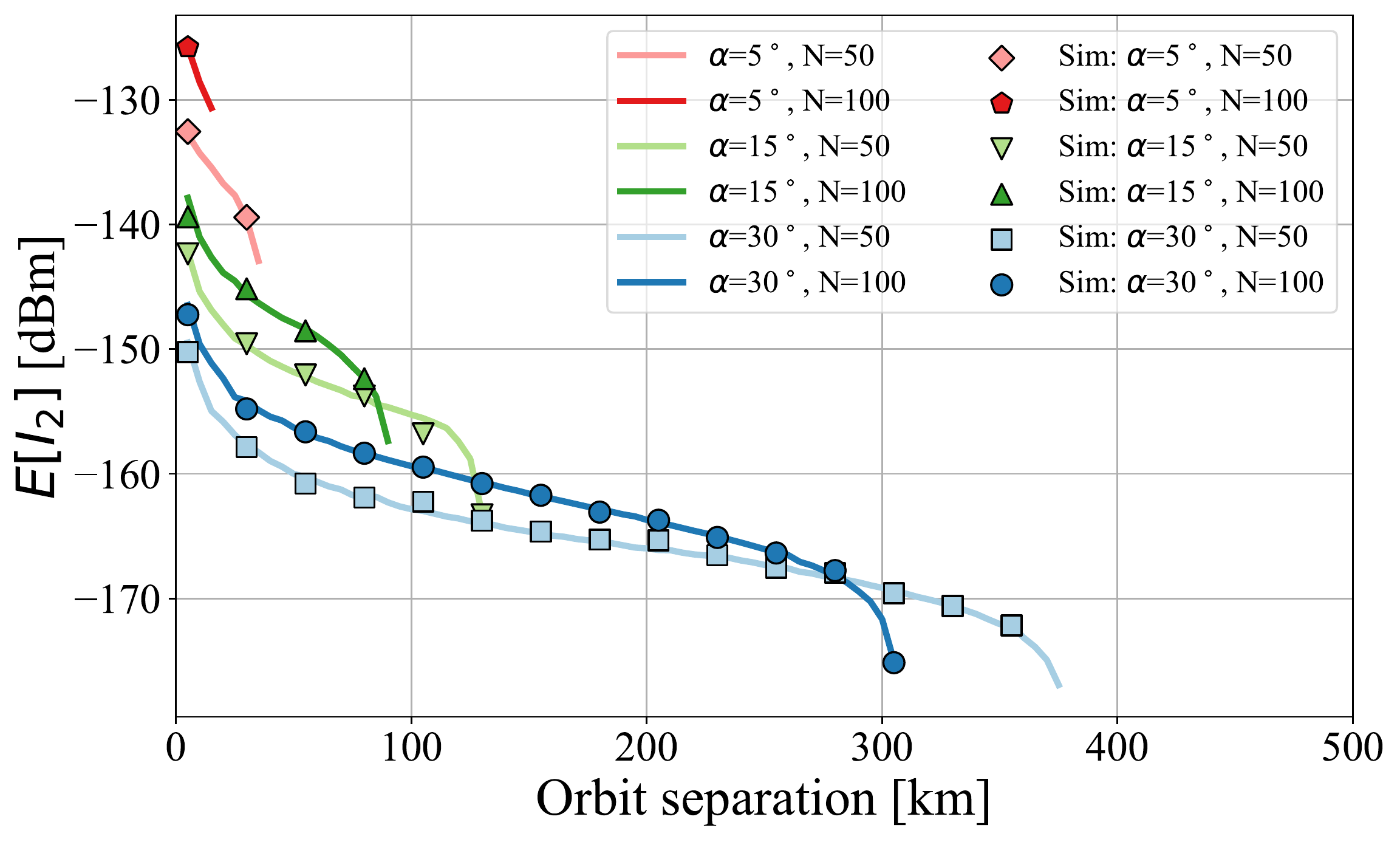}
        \label{fig:coplanar_interference_altitude_results}
    }
    \captionsetup{justification=centering}
    \vspace{-1mm}
    \caption{Average interference and SIR for co-planar orbits.}
    \label{fig:coplanar_results}
\end{figure*}

\subsection{Single orbit}
\label{sec:single_res}
We start with Fig.~\ref{fig:single_orbit_results} presenting the average SIR for the single orbit setup, $S_{1}$, as a function of the number of satellites in the orbit, $N$. First, while the altitude for this figure is set to $500$\,km, it has been observed in (\ref{eq:s1}) and later confirmed with simulations that the orbit altitude, $h$, does not impact the average SIR, as the altitude-dependent parameters cancel each other. Then, Fig.~\ref{fig:single_orbit_results} illustrates that for a fixed beamwidth, $\alpha$, a higher number of satellites, in general, leads to greater interference and, consequently, lower average SIR. However, this effect is non-monotonic. Particularly, there are specific \emph{drop} points in Fig.~\ref{fig:single_orbit_results}, where the SIR notably decreases from $k$ to $k+1$ satellites. These are the points when the number of interfering satellites, $N_{i}$, gets incremented.

Notably, the impact of every new interfering satellite on the average SIR is smaller and smaller, i.e., for $\alpha=30\degree$ the drop in SIR from 24 to 25 satellites is $>$$1.5$\,dB, while the corresponding SIR decrease from 95 to 96 satellites is $<$$0.2$\,dB. We also observe that the intuitive dependency of SIR increasing with antenna directivity (demonstrated in~\cite{HeathInterference} and~\cite{Petrov2017interference}, among other studies) holds for satellite deployments as well. Nevertheless, even for narrow-beam links of $5\degree$ only, the SIR can be as low as $6$\,dB, thus challenging the reliability and performance of the cross-link if not properly accounted for. We finally notice a close-to-perfect match between the analytical curves and the performed Monte-Carlo simulations.

\subsection{Co-planar orbits}
\label{sec:coplanar_res}
We then study the interference from a co-planar orbit,~$E[I_2]$.

\subsubsection{Time-dependent interference}
Due to non-stationary mutual positions and orientations of the satellites in different orbits, the average SIR, $S_{2}$, evolves in time. This effect is illustrated in Fig.~\ref{fig:coplanar_SIR_time_results}, modeling two co-planar orbits with $N=N_c$ and the altitudes of $500$\,km and $510$\,km (hence, with only $10$\,km separation distance). Despite a relatively small separation distance between the orbits, the average SIR fluctuates between $5$\,dB and $-30$\,dB by following a periodic pattern. We observe that the period depends on the number of satellites in the orbits, particularly becoming shorter with more satellites. This is because the higher the number of satellites per orbit, the lower the angular separation between them, and the rotation required to achieve a redundant satellite relative position is longer. Additionally, we again observe that the beamwidth plays a critical role, with the curves for wider beamwidth always staying below 0~dB (interference is higher than the received signal). These results, however, are influenced by the close separation between orbits.
\subsubsection{Time-averaged interference}
We proceed with exploring the impact of the separation distance between the orbits (the difference in their attitudes) on the time-averaged SIR, as illustrated in Fig.~\ref{fig:coplanar_interference_altitude_results}. Here, we first notice a certain separation distance, starting from which the satellites do not interfere with each other anymore. This is an important practical outcome from our study, allowing, i.e., full frequency reuse among the satellite constellations deployed with an altitude difference greater than $X$.

\begin{figure*}[!t]
  \centering
  \subfigure[SIR evolution in time]
  {
    \includegraphics[width=0.315\textwidth]{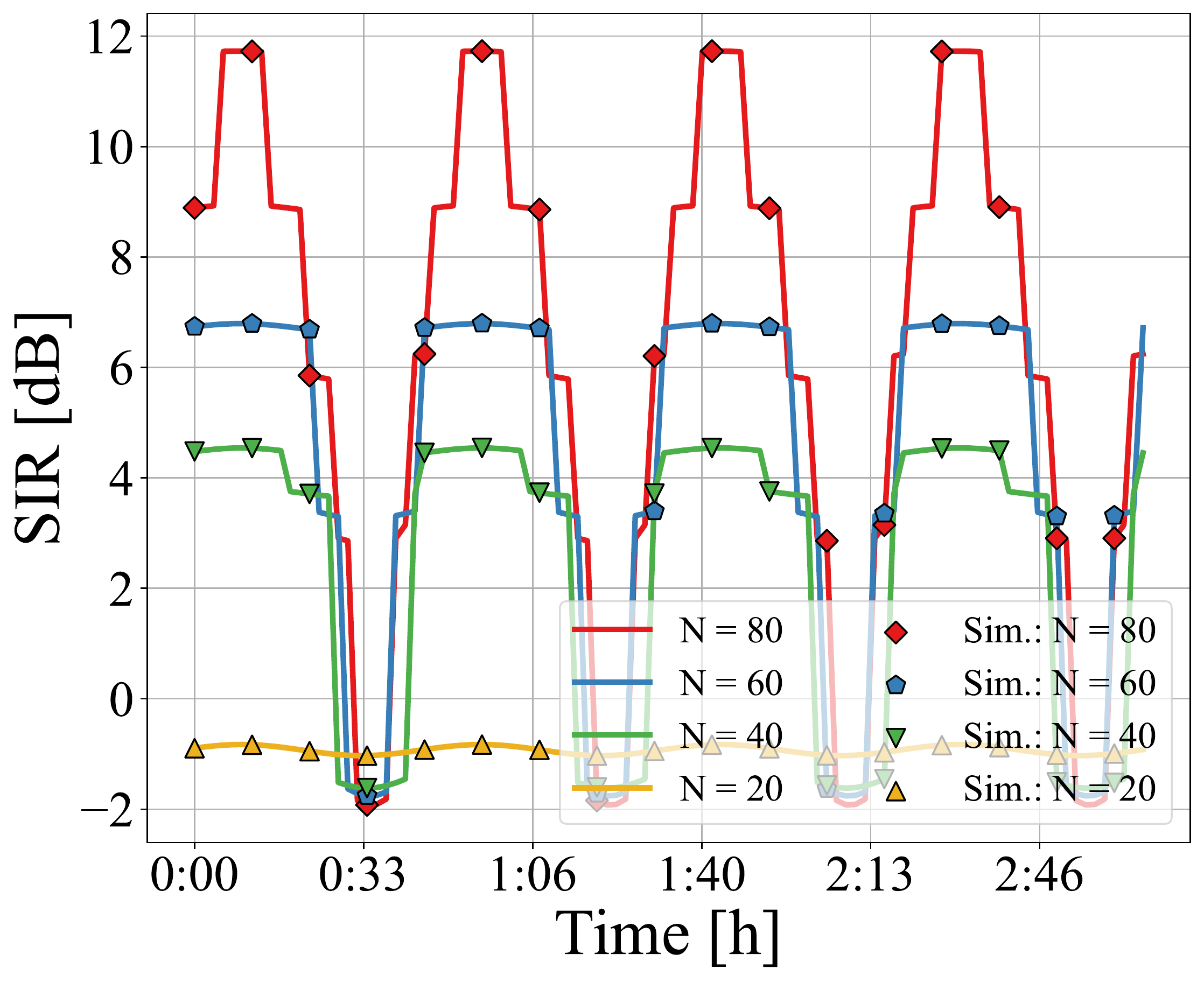}
    \label{fig:shifted_time_results}
     
  }
  \hfill
  \subfigure[The effect of beamwidth]
  {
    \includegraphics[width=0.315\textwidth]{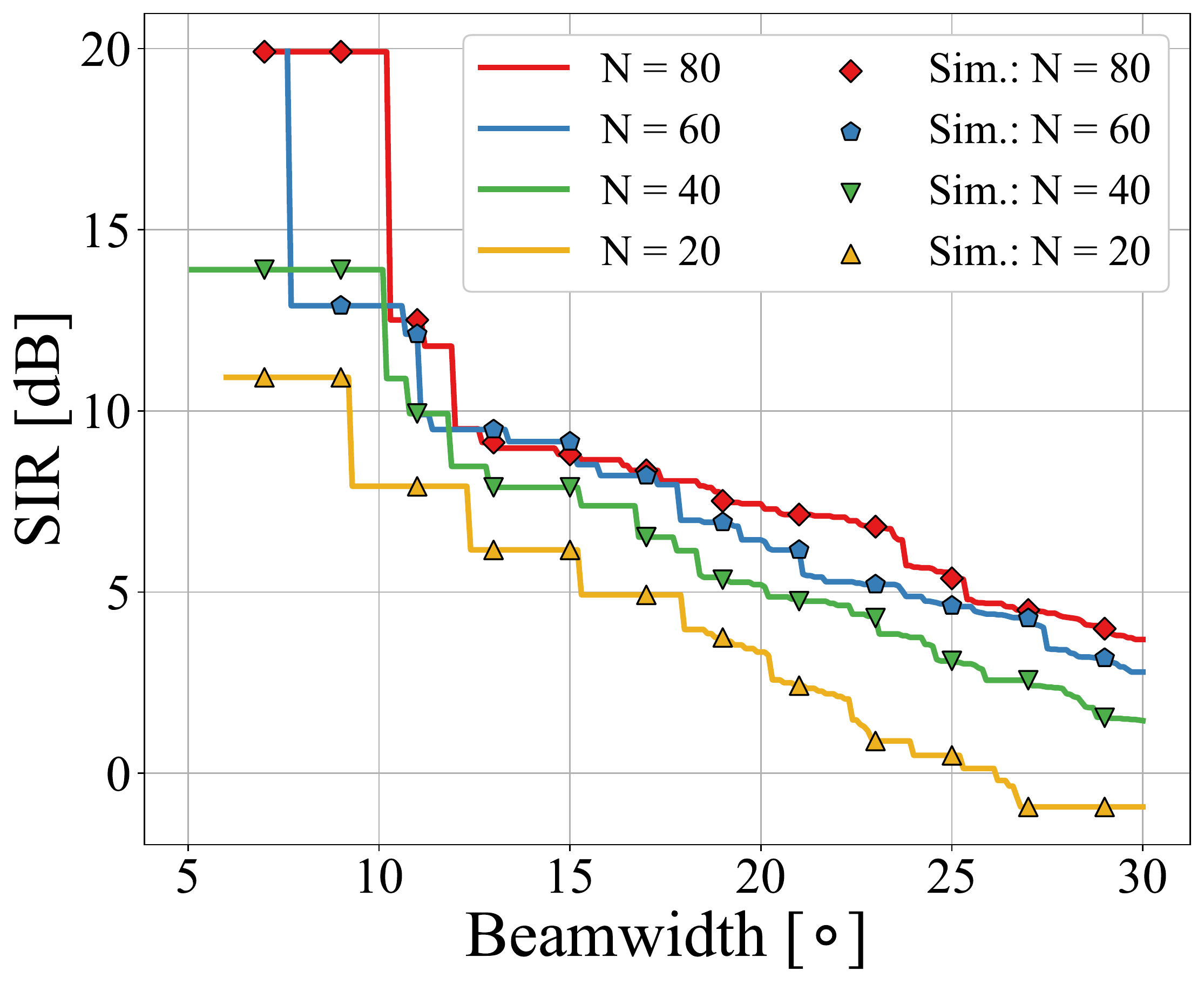}
    \label{fig:shifted_beamwidth_results}
     
  }
  \hfill
  \subfigure[The effect of inclination]
  {
    \includegraphics[width=0.315\textwidth]{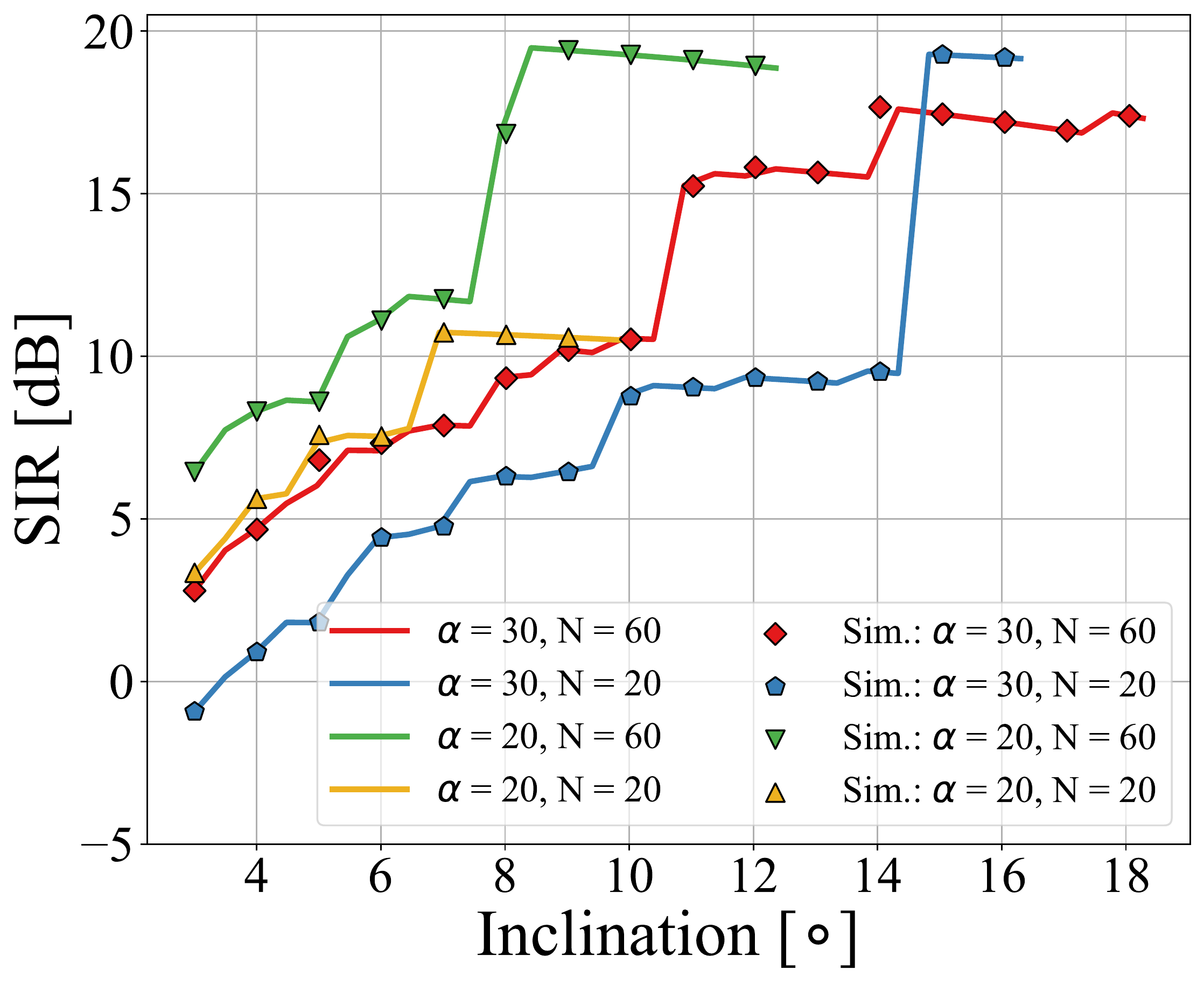}
    \label{fig:shifted_inclination_results}
  }
  \captionsetup{justification=centering}
  \vspace{-2mm}
  \caption{SIR for the interference coming from a shifted orbit.}
  \label{fig:shifted_results}
\end{figure*}

The second key observation from Fig.\ref{fig:coplanar_interference_altitude_results} is the fact that the interference from a co-planar orbit is higher for a greater number of satellites, but drops faster after a certain orbit separation leading to the crossing of the corresponding curves. This effect is caused by the fact that the satellites pointing directions depend on the number of satellites in orbit. As the satellites communicate with their direct neighbors, a larger number of satellites in a co-planar orbit leads to more interferers but also steers interfering signals away from the Rx. At a certain orbit separation, the latter effect starts dominating the former one.

\subsection{Shifted orbits}\label{sec:shifted_res}
The last part of the numerical study focuses on the interference coming from a shifted orbit, as explored in Fig.~\ref{fig:shifted_results}. For comparison, we present the results in all the subfigures for the same altitude of $500$\,Km and the same relative offset of $0^{\circ}$.

\subsubsection{Time-dependent interference}
Similar to a co-planar setup, the impact of the interference is time-variant, as in Fig.~\ref{fig:shifted_time_results}. We illustrate the setup, where an inclination is $3^{\circ}$, $N$$=$$N_s$, and $\alpha=30^{\circ}$. The time duration of the results corresponds to two orbital periods. Again we observe periodicity, this time corresponding to the orbital period, with two clear SIR drops per period (four in total in the figure). These drops correspond to the moments when the two orbits cross.

\subsubsection{Time-averaged interference}

We then proceed with Fig.~\ref{fig:shifted_beamwidth_results}, presenting the time-averaged SIR, $S_{3}$, as a function of beamwidth, $a$. We first observe that there is a ``threshold'' beamwidth from which interference starts to appear (for smaller $\alpha$, SIR goes to infinity). For $3^{\circ}$ inclination, this threshold is $\alpha$$\approx$$6^{\circ}$. We also observe a step-like decreasing behavior of the curves, meaning that larger beamwidths result in a reduction of SIR. Hence, from this perspective, the interference from a shifted orbit behaves similarly to the one from the same orbit, as in Fig.~\ref{fig:single_orbit_results}.

The results also notably depend on the orbit inclination angle, as explored in Fig.~\ref{fig:shifted_inclination_results}. Here, higher inclination angles, in general, yield better SIR values. Certain thresholds are again present, after which there is no interference (around $\gamma=9^{\circ}$ to for $\alpha=20^{\circ}$ and $\gamma=18^{\circ}$ for $\alpha=30^{\circ}$, respectively). We also observe that there are specific small regions with an opposite trend, i.e., $\gamma=14^{\circ}$ to $\gamma=17^{\circ}$ for $\alpha=30^{\circ}$ and $N=60$. This effect is primarily due to the fact that in these inclination regions, the particular satellites interfering at the moment get closer to the Rx, while still within the Rx beam, thus decreasing the average SIR. When the inclination grows further (i.e., over $17^{\circ}$), some potential interferers are still close to the Rx, but not anymore within the Rx beam. So, the SIR keeps growing further. The latter is a useful finding for practical constellations design, allowing to reduce the impact of interference even with a high number of satellites per orbit.

\section{Conclusions}
\label{sec:conclusions}
Directional cross-links over mmWave and, in the future, THz bands are important technology enablers for high-rate LEO satellite connectivity services as an integral part of 5G-Advanced and 6G networks. In this paper, we have presented a novel mathematical framework to model the interference among these cross-links in selected prospective scenarios. We have also complemented our mathematical analysis with an extensive simulation campaign, identifying both temporal and time-averaged interference-related characteristics. The delivered models can be extended further by accounting for additional orbits, such as polar orbits, imperfect beam alignment, and more sophisticated specific antenna radiation patterns.

Our numerical results indicate that, despite the use of directional antennas, cross-link interference can be non-negligible in some of the possible deployment configurations. The observed effect may challenge the reliability and performance of cross-links within the single LEO satellite constellation, as well as the efficient co-existence of neighboring satellite constellations. Hence, the interference among mmWave and THz cross-links cannot be always neglected, but rather should be explored further and accounted for in the design and deployment of next-generation LEO satellite networks.

\balance
\section*{Acknowledgment}
This work has been supported in part by the projects CNS-1955004 and CNS-2011411 by the National Science Foundation (NSF), as well as by the U.S. Air Force FA8750-20-1-0200. This project was also supported in part by a fellowship from ``la Caixa'' foundation (ID 100010434, LCF/BQ/AA20/11820041).

\vspace{7mm}
\bibliography{main}

\newcommand{\noopsort}[1]{} \newcommand{\printfirst}[2]{#1}
  \newcommand{\singleletter}[1]{#1} \newcommand{\switchargs}[2]{#2#1}
\begin{thebibliography}{10}
\providecommand{\url}[1]{#1}
\csname url@samestyle\endcsname
\providecommand{\newblock}{\relax}
\providecommand{\bibinfo}[2]{#2}
\providecommand{\BIBentrySTDinterwordspacing}{\spaceskip=0pt\relax}
\providecommand{\BIBentryALTinterwordstretchfactor}{4}
\providecommand{\BIBentryALTinterwordspacing}{\spaceskip=\fontdimen2\font plus
\BIBentryALTinterwordstretchfactor\fontdimen3\font minus
  \fontdimen4\font\relax}
\providecommand{\BIBforeignlanguage}[2]{{%
\expandafter\ifx\csname l@#1\endcsname\relax
\typeout{** WARNING: IEEEtran.bst: No hyphenation pattern has been}%
\typeout{** loaded for the language `#1'. Using the pattern for}%
\typeout{** the default language instead.}%
\else
\language=\csname l@#1\endcsname
\fi
#2}}
\providecommand{\BIBdecl}{\relax}
\BIBdecl

\bibitem{Giordani2021}
M.~Giordani and M.~Zorzi, ``Non-terrestrial networks in the {6G} era:
  {C}hallenges and opportunities,'' \emph{IEEE Network}, vol.~35, no.~2, pp.
  244--251, March/April 2021.

\bibitem{6g_LEO_commag}
X.~Lin \emph{et~al.}, ``On the path to {6G}: Embracing the next wave of low
  earth orbit satellite access,'' \emph{IEEE Communications Magazine}, vol.~59,
  no.~12, pp. 36--42, Dec 2021.

\bibitem{Zhu2022IoT}
X.~Zhu and C.~Jiang, ``Integrated satellite-terrestrial networks toward {6G}:
  {A}rchitectures, applications, and challenges,'' \emph{IEEE Internet of
  Things Journal}, vol.~9, no.~1, pp. 437--461, Jan 2022.

\bibitem{pachler21}
N.~Pachler, I.~del Portillo, E.~F. Crawley, and B.~G. Cameron, ``An updated
  comparison of four low earth orbit satellite constellation systems to provide
  global broadband,'' in \emph{Proc. of the IEEE ICC Workshops}, June 2021, pp.
  1--7.

\bibitem{OzgurTHzSpace}
M.~Civas and O.~B. Akan, ``Terahertz wireless communications in space,''
  \emph{ITU J. on Future and Evolving Tech.}, vol.~2, no.~7, pp. 31--38, Oct
  2021.

\bibitem{aliaga2022joint}
S.~Aliaga, A.~J. Alqaraghuli, and J.~M. Jornet, ``Joint terahertz communication
  and atmospheric sensing in low earth orbit satellite networks: Physical layer
  design,'' in \emph{Proc. of the IEEE WoWMoM}, June 2022, pp. 457--463.

\bibitem{Chaudhary2022hybrid}
S.~Chaudhary, A.~Sharma, and N.~Chaudhary, ``6 × 20 {Gbps} hybrid wdm–pi
  inter-satellite system under the influence of transmitting pointing errors,''
  \emph{Journal of Optical Communications}, vol.~37, no.~4, pp. 375--379, May
  2016.

\bibitem{Carlson2022}
R.~T. Carlson, ``Architecture for reconfigurable next-generation lasercom
  terminals,'' in \emph{Proc. of the IEEE ICSOS}, March 2022, pp. 203--212.

\bibitem{Akyildiz_OLD}
I.~F. Akyildiz, C.~Han, Z.~Hu, S.~Nie, and J.~M. Jornet, ``Terahertz band
  communication: {A}n old problem revisited and research directions for the
  next decade,'' \emph{IEEE Transactions on Communications}, vol.~70, no.~6,
  pp. 4250--4285, June 2022.

\bibitem{Roy2021Spaceborne}
R.~J. Roy, M.~Lebsock, and M.~J. Kurowski,
  ``\BIBforeignlanguage{English}{Spaceborne differential absorption radar water
  vapor retrieval capabilities in tropical and subtropical boundary layer cloud
  regimes},'' \emph{\BIBforeignlanguage{English}{Atmospheric Measurement
  Techniques}}, vol.~14, no.~10, p. 6443–6468, Oct 2021.

\bibitem{Brown2033Fundamentals}
E.~R. Brown, ``Fundamentals of terrestrial millimeter-wave and {THz} remote
  sensing,'' \emph{International Journal of High Speed Electronics and
  Systems}, vol.~13, no.~04, pp. 995--1097, Dec 2003.

\bibitem{ieee_standard_thz_m}
V.~Petrov, T.~Kurner, and I.~Hosako, ``{IEEE 802.15.3d: First standardization
  efforts for sub-terahertz band communications toward 6G},'' \emph{IEEE
  Communications Magazine}, vol.~58, no.~11, pp. 28--33, Nov 2020.

\bibitem{Sen2022multi}
P.~Sen, J.~V. Siles, N.~Thawdar, and J.~M. Jornet,
  ``\BIBforeignlanguage{en}{Multi-kilometre and multi-gigabit-per-second
  sub-terahertz communications for wireless backhaul applications},''
  \emph{\BIBforeignlanguage{en}{Nature Electronics}}, p. 1–12, Dec 2022.

\bibitem{joonas_thz_airplane}
J.~Kokkoniemi, J.~M. Jornet, V.~Petrov, Y.~Koucheryavy, and M.~Juntti,
  ``Channel modeling and performance analysis of airplane-satellite terahertz
  band communications,'' \emph{IEEE Transactions on Vehicular Technology},
  vol.~70, no.~3, pp. 2047--2061, March 2021.

\bibitem{PengWC2022}
D.~Peng, D.~He, Y.~Li, and Z.~Wang, ``Integrating terrestrial and satellite
  multibeam systems toward {6G}: {T}echniques and challenges for interference
  mitigation,'' \emph{IEEE Wireless Communications}, vol.~29, no.~1, pp.
  24--31, Feb 2022.

\bibitem{RappaportCL2021}
Y.~Xing and T.~S. Rappaport, ``Terahertz wireless communications: {C}o-sharing
  for terrestrial and satellite systems above 100 {GHz},'' \emph{IEEE
  Communications Letters}, vol.~25, no.~10, pp. 3156--3160, Oct 2021.

\bibitem{Springer2021}
Z.~Lin, J.~Jin, J.~Yan, and L.~Kuang, ``A method for calculating the
  probability distribution of interference involving mega-constellations,''
  \emph{Adv. in Astronautics Sc. and Tech.}, vol.~4, pp. 107--117, June 2021.

\bibitem{KaragiannidisJSAC2022}
Y.~Zhang, H.~Zhang, H.~Zhou, K.~Long, and G.~K. Karagiannidis, ``Resource
  allocation in terrestrial-satellite-based next generation multiple access
  networks with interference cooperation,'' \emph{IEEE Journal on Selected
  Areas in Commun.}, vol.~40, no.~4, pp. 1210--1221, April 2022.

\bibitem{Polese2022}
M.~Polese \emph{et~al.}, ``Dynamic spectrum sharing between active and passive
  users above 100 {GHz},'' \emph{Nature Communications Engineering}, vol.~1,
  no.~11, p. 1–9, May 2022.

\bibitem{Mendoza2017}
H.~A. Mendoza, G.~Corral-Briones, J.~M. Ayarde, and G.~G. Riva, ``Spectrum
  coexistence of {LEO} and {GSO} networks: {A}n interference-based design
  criteria for {LEO} inter-satellite links,'' in \emph{Proc. of the CLEI}, Sept
  2017, pp. 1--6.

\bibitem{MayorgaTWC2021}
I.~Leyva-Mayorga, B.~Soret, and P.~Popovski, ``Inter-plane inter-satellite
  connectivity in dense {LEO} constellations,'' \emph{IEEE Transactions on
  Wireless Communications}, vol.~20, no.~6, pp. 3430--3443, June 2021.

\bibitem{MayorgaGlobecom2011}
I.~Leyva-Mayorga, M.~Röper, B.~Matthiesen, A.~Dekorsy, P.~Popovski, and
  B.~Soret, ``Inter-plane inter-satellite connectivity in {LEO} constellations:
  {B}eam switching vs. beam steering,'' in \emph{Proc. of the IEEE GLOBECOM},
  Dec 2021, pp. 1--6.

\bibitem{HeathInterference}
K.~Venugopal, M.~C. Valenti, and R.~W. Heath, ``Device-to-device millimeter
  wave communications: Interference, coverage, rate, and finite topologies,''
  \emph{IEEE Transactions on Wireless Communications}, vol.~15, no.~9, pp.
  6175--6188, Sept 2016.

\bibitem{Petrov2017interference}
V.~Petrov, M.~Komarov, D.~Moltchanov, J.~M. Jornet, and Y.~Koucheryavy,
  ``Interference and {SINR} in millimeter wave and terahertz communication
  systems with blocking and directional antennas,'' \emph{IEEE Trans. on
  Wireless Commun.}, vol.~16, no.~3, p. 1791–1808, Mar 2017.

\bibitem{Leyva2022ngso}
I.~Leyva-Mayorga, B.~Soret, B.~Matthiesen, M.~Röper, D.~Wübben, A.~Dekorsy,
  and P.~Popovski, ``{NGSO} constellation design for global connectivity,'' no.
  arXiv:2203.16597, April 2022.

\bibitem{Maral2009satellite}
G.~Maral, M.~Bousquet, and Z.~Sun, \emph{\BIBforeignlanguage{en}{Satellite
  communications systems: {S}ystems, techniques and technology}}, 5th~ed.\hskip
  1em plus 0.5em minus 0.4em\relax John Wiley, 2009.

\bibitem{Sharaf2012satellite}
M.~A.~Sharaf, ``\BIBforeignlanguage{en}{Satellite to satellite visibility},''
  \emph{\BIBforeignlanguage{en}{The Open Astronomy Journal}}, vol.~5, no.~1, p.
  26–40, May 2012.

\bibitem{Alqaraghuli2021compact}
A.~J. Alqaraghuli, A.~Singh, and J.~M. Jornet, ``Compact high-gain dual-band
  antenna for full-duplex terahertz communication in cubesat
  mega-constellations,'' in \emph{Proc. of the IEEE APS/URSI}, 2021, pp.
  1827--1828.

\end{thebibliography}
\bibliographystyle{IEEEtran}
\end{document}